\documentclass[a4paper]{article}

\usepackage{amsfonts}

\def\be{\begin{equation}}
\def\ee{\end{equation}}
\def\bea{\begin{eqnarray}}
\def\eea{\end{eqnarray}}
\def\({\left(}
\def\){\right)}
\def\<{\left<}
\def\>{\right>}

\def\[{\left[}
\def\]{\right]}

\def\be{\begin{equation}}
\def\ee{\end{equation}}
\def\bea{\begin{eqnarray}}
\def\eea{\end{eqnarray}}
\def\({\left(}
\def\){\right)}
\def\<{\left<}
\def\>{\right>}

\def\[{\left[}
\def\]{\right]}

\def\+{\bar}

\def\Tr{{\mbox{Tr}}}

\def\L{{\cal{L}}}

\def\Ordo{{\cal{O}}}

  \usepackage{feynmf}
  \begin{document}\setlength{\unitlength}{1mm}

\pagestyle{empty}
\vskip-10pt
\vskip-10pt
\hfill 
\begin{center}
\vskip 3truecm
{\Large \bf
One-loop corrections to Bagger-Lambert theory}\\ 
\vskip 2truecm
{\large \bf
Andreas Gustavsson\footnote{a.r.gustavsson@swipnet.se}}\\
\vskip 1truecm
{\it F\"{o}rstamajgatan 24,\\
S-415 10 
G\"{o}teborg, Sweden}\\
\end{center}
\vskip 2truecm
{\abstract{We rewrite the Bagger-Lambert action for any Lie 3-algebra as a standard Chern-Simons action coupled to matter. We use this action to compute self-energies and vertex corrections at one-loop order. Non-renormalization of the coupling constant comes out as a direct consequence of the Lie 3-algebra structure underlying the Lie algebra.}}

\vfill 
\vskip4pt
\eject
\pagestyle{plain}

\section{Introduction}
Maximally supersymmetric theories in $1+2$ dimensions with $SO(8)$ R-symmetry were found in \cite{Bagger:2007jr, Gustavsson:2007vu}. One expects these theories to describe a field theory on parallel M2 branes. In \cite{Bandres:2008vf} it was shown that the theory is conformally invariant, at least at the classical level. One feature of these theories is that the matter fields take values in a Lie 3-algebra. If the generators are denoted $T^a$, then a Lie 3-algebra ${\cal{A}}$ is defined by 
\bea
[T^a,T^b,T^c] &=& f^{abc}{}_d T^d.
\eea
where the structure constants $f^{abc}{}_d$ are totally antisymmetric in $a,b,c$ and are subject to the fundamental identity, 
\bea
f^{[abc}{}_g f^{d]eg}{}_h = 0
\eea
which resembles the Jacobi identity for Lie algebras. 

The gauge field takes values in the Lie algebra associated with the Lie 3-algebra. Hence this is usual a gauge theory, and the Lie 3-algebra is just an additional restriction that we put on this gauge theory. This additional restriction is required from supersymmetry and has persisted all attempts of weakening.

In \cite{Papadopoulos:2008gh, Gauntlett:2008uf} it was proven that the only finite-dimensional solution to a Lie 3-algebra are the ones associated with $SO(4)$ Lie algebra, if one assumes a positive Killing form on the Lie 3-algebra. To describe $N=1,2,..$ parallel M2's one would expect to have a Lie 3-algebra associated to each $N$. In \cite{Gran:2008vi} this was partially achieved by ignoring the Killing form on the Lie 3-algebra. In \cite{Ho:2008ei, Benvenuti:2008bt, Gomis:2008uv} (see also \cite{FigueroaO'Farrill:2008zm,Lin:2008qp}) the Killing form for a closely related class of Lie 3-algebras was obtained. This Killing form has one negative eigenvalue and the associated theories where found to have no coupling constant at all.

The Bagger-Lambert action involves a Chern-Simons action. The Chern-Simons action is not totally gauge invariant, but changes by $2\pi$ times an integer under large gauge transformations \cite{Witten:1988hf}. Hence, for this action to make sense, the gauge coupling constant must not recieve any (non-integer) quantum corrections. Apriori we can imagine different scenarios. It could be that the coupling does not renormalize for any choice of Lie 3-algebra. Or it could be that we find some additional constraints on possible Lie 3-algebras that yield consistent quantum theories, or there could be no Lie 3-algebras yielding a consistent quantum theory. In this paper we will show some indicatation that it may be that any Lie 3-algebra yields a consistent theory. 

For the theories found in \cite{Ho:2008ei, Benvenuti:2008bt,Gomis:2008uv} the coupling constant can be absorbed by a field redefinition. But the action in these papers is yet nothing but a rewriting of the Bagger-Lambert action for a particular choice of associated Lie algebra. Hence one would at first sight suspect the action not being completely gauge invariant, but would change as any Chern-Simons action does, under large gauge transformations. Then it appears that also these theories would have a discreteness value of the coupling constant, which then can be put equal to one by a field redefintion. Being then a strongly coupled theory we should seek some other parameter which we can take small if we want to study the quantum theory using a perturbation expansion. 

Also there is an infinite class of infinite-dimensional solutions to the fundamental identity \cite{Ho:2008nn} that could be a physical relevance in the large $N$ limit.

In this paper we will therefore make no assumptions of the Lie 3-algebra. We will compute one-loop quantum corrections. Previous computations of this type has been carried out in \cite{Chen:1992ee} for Chern-Simons gauge theory coupled to matter fields. 

In \cite{Kapustin:1994mt} it was shown that the beta function associated to the gauge coupling constant vanishes for a Chern-Simons theory coupled matter fields if no relations are assumed between the various coupling constants. In this paper we impose relations between the coupling constants in the classical action, in such a way that there is just one free parameter in the theory. This implies tougher consistency constraints (Ward identities) on the quantum theory and consequently it is no longer apparent that any such theory would be quantum mechanically consistent. If one assumes that all coupling constants are linearly dependent and there is only one free parameter, then it seems likely that quantum consistency alone requires the theory be highly supersymmetric -- hence most classical Chern-Simons-matter actions one can write down which have just one free parameter, are quantum mechanically inconsistent, and only highly supersymmetric actions have any chance of giving a consistent quantum theory. Of course this is no longer the case if one allows more freedom among the coupling constants \cite{Kapustin:1994mt}. In this paper we make a quite general ansatz for a Chern-Simons-matter theory (albeit not the most general ansatz). We then show that within our ansatz, the only quantum mechanically consistent theory is the one with $N=8$ supersymmetry constructed in \cite{Bagger:2007jr, Gustavsson:2007vu}. 

One can also run the argument in the other direction. By assuming the quantum theory is supersymmetric one can deduce from general argument that $N=2$ and $N=3$ supersymmetric Chern-Simons-matter theory are quantum mechanically consistent \cite{Gaiotto:2007qi}. The $N=3$ theories are particularly interesting because for particular choices of gauge group and matter field representations one can get $N=6$ supersymmetry \cite{Aharony:2008ug}. Moreover restricting the gauge group to $SO(4)$ one finds $N=8$ supersymmetry. Since $N=3$ theory is consistent, so must also $N=8$ theory with $SO(4)$ gauge group be consistent.

In section \ref{lie} we carefully discuss Lie algebras associated with Lie 3-algebras, and obtain relations between various Casimir invariants for such Lie algebras. In section \ref{action} we rewrite the Bagger-Lambert action as a normal gauge theory. In remaining sections we compute one-loop diagrams and find non-renormalization of the coupling constant as a consequence of the fundamental identity.

\section{Associated Lie algebras of Lie 3-algebras}\label{lie}
The fundamental identity
\bea
f^{[abc}_g f^{d]eg}{}_f &=& 0\label{v1}
\eea
is equivalent with the identity
\bea
f^{abc}{}_g f^{deg}{}_f &=& 3 f^{de[a}{}_{g} f^{bc]g}{}_f.\label{v2}
\eea
This equivalence was proven by Gran \cite{Gran:2008vi}. The proof by Gran goes as follows. First assume that Eq (\ref{v1}) holds. We can write this equation as
\bea
f^{abc}{}_g f^{deg}{}_f &=& 3 f^{d[ab}{}_g f^{c]eg}{}_f\label{v3}
\eea
We note that the right-hand side must be antisymmetric in $d,e$ whenever Eq (\ref{v1}) holds (which it does by our assumption) simply because the left-hand side is antisymmetric in $d,e$. Applying Eq (\ref{v1}) once again on the right-hand side, now by antisymmetrizing in $a,b,d,e$ instead, we get
\bea
f^{d[ab}{}_g f^{c]eg}{}_f - f^{e[ab}{}_g f^{c]dg}{}_f = 2 f^{de[a}{}_g f^{bc]g}{}_f
\eea
But now, remembering the aforementioned antisymmetry in $d,e$, we can write this as just
\bea
f^{d[ab}{}_g f^{c]eg}{}_f = f^{de[a}{}_g f^{bc]g}{}_f.
\eea
Substituting this back into the right-hand side of Eq (\ref{v3}), we get Eq (\ref{v2}). The converse is shown in a similar (and perhaps even simpler) way: assuming Eq (\ref{v2}) we can derive Eq (\ref{v1}) by applying Eq (\ref{v2}) twice.

The fundamental identity in the form of Eq (\ref{v2}) can also be written as \cite{Gustavsson:2008dy}
\bea
[t^{ab},t^{cd}](T^e) &=& -2f^{ab[c}{}_g t^{d]g}(T^e)
\eea
where we define linear maps
\bea
t^{ab} = [T^a,T^b,\bullet]
\eea
Let us denote by ${\cal{G}}$ the set of such linear maps acting on basis elements $T^a$ of the Lie 3-algebra ${\cal{A}}$. Then the above shows that the commutator of two elements in ${\cal{G}}$ is again an element in ${\cal{G}}$. This is suggestive of a Lie algebra, where the generators act in the fundamental representation as
\bea
(t^{ab})^{c}{}_d = f^{abc}{}_d.\label{rep}
\eea
If so, then we read off its structure constants from the commutator as
\bea
C^{ab,cd}{}_{ef} &=& 2f^{ab[c}{}_{[e} \delta^{d]}_{f]} \label{structure1}
\eea
Two immediate questions now arise. First, are these structure constants antisymmetric in the pair of indices $ab$ and $cd$? Second, do these structure constants satisfy the Jacobi identity
\bea
C^{ab,cd}{}_{ef} C^{gh,ef}{}_{kl} + C^{gh,ab}{}_{ef} C^{cd,ef}{}_{kl} +C^{cd,gh}{}_{ef} C^{ab,gh}{}_{kl} &=& 0\label{Jac}
\eea
of a Lie algebra? 

At first sight the antisymmetry in $ab$ and $cd$ looks impossible, and seems not to follow from the fundamental identity in any way. But let us now also assume there is a Killing form $h^{ab}$ on ${\cal{A}}$. This then can be used to get a completely antisymmetric tensor $f^{abcd}=f^{abc}{}_e h^{ed}$. Moreover this tensor is invariant under the action of the generators $t^{gh}$ in the fundamental representation Eq (\ref{rep}) as a direct consequence of the fundamental identity. This is suggestive of a Killing form $\kappa^{ab,cd} = f^{abcd}$ of a Lie algebra. Let us denote by $\kappa_{ab,cd}$ its inverse, $\kappa_{ab,cd}\kappa^{cd,ef} = \delta_{ab}^{ef}$. Using this we can now reshape the fundamental identity as the linear equation\footnote{One may wonder how we can write the fundamental identity as a {\sl{linear}} equation in the structure constants $f$. Rewriting the ordinary Jacobi identity as a linear equation is impossible for Lie algebras. The reason to that is that for Lie algebras we have a Killing form which must be different from the structure constants. Here the Killing form of ${\cal{G}}$ can be taken to be the same as the structure constants of ${\cal{A}}$ and this enable us to write the fundamental identity as a linear equation.}
\bea
f^{[abc}{}_g \delta^{d]g}_{eh} = 0.
\eea
which is nothing but the sought for antisymmetry property
\bea
f^{ab[c}{}_{[h} \delta^{d]}_{e]} + f^{cd[a}{}_{[h} \delta^{b]}_{e]} = 0.
\eea
As an example this may be checked for the case of $SO(4)$ where we have  
\bea
f^{abcd} &=& \epsilon^{abcd}\cr
h^{ab} &=& \delta^{ab}
\eea
We then have for instance
\bea
\epsilon^{12[3}{}_{[4} \delta^{2]}_{2]} = - \epsilon^{32[1}{}_{[4} \delta^{2]}_{2]}.
\eea

To settle the second question, we should sum the terms\footnote{Here $T^{[a|bcd|e]}$ means $(T^{abcde} - T^{ebcda})/2$.}
\bea
C^{ab,cd}{}_{gh} C^{ef,gh}{}_{mn} &=& 4f^{ab[c}{}_{k} f^{|efk|}{}_{[m} \delta^{d]}_{n]}\cr
C^{cd,ef}{}_{gh} C^{ab,gh}{}_{mn} &=& - C^{ef,cd}_{gh} C^{ab,gh}{}_{mn} = -4f^{ef[c}{}_g f^{abg}{}_{[m} \delta^{d]}_{n]}\cr
C^{ef,ab}{}_{gh} C^{cd,gh}{}_{mn} &=& 4f^{ef[a}{}_g f^{|cdg|}{}_{[m} \delta^{b]}_{n]} = -4f^{ef[a}{}_g f^{b]cg}{}_{[m} \delta^d_{n]} - ...
\eea
In the last line we used the fundamental identity in the form $f^{[cdg}{}_{[m} \delta^{b]}_{n]} = 0$. Collecting all terms associated with $\delta^d_m$ we now find the coefficient
\bea
f^{abc}{}_k f^{efk}{}_m - f^{efc}{}_g f^{abg}{}_m - 2 f^{ef[a}_g f^{b]cg}{}_m
\eea
and this vanishes identically, being the fundamental identity in its original form. By symmetry all terms in the sum vanish, there being nothing particular with the choice $\delta^d_n$.

Quite generally we can consider two Killing forms on the Lie algebra associated with a Lie 3-algebra. Assume there is a Killing form $h^{ab}$ on the Lie 3-algebra, by which is meant a tensor subject to the invariance condition
\bea
f^{abc}{}_e h^{ed} + f^{bcd}{}_e h^{ae} &=& 0.
\eea
This condition can be read in three different ways. First it says that $f^{abcd} = f^{abc}{}_e h^{ed}$ is totally antisymmetric. Second, it says that $h^{ab}$ is an invariant tensor in the associated Lie algebra generated by $(t^{ab})^c{}_d = f^{abc}{}_d$. And third, it says that $h^{ab}$ is a central element, commuting with any Lie algebra generator $t^{ab}$.
Given such a Killing form, we may consider two invariant tensors of the required structure of a Killing form on the associated Lie algebra, namely
\bea
\kappa^{ab,cd} &=& f^{abcd}\\
g^{ab,cd} &=& f^{abe}{}_f f^{cdf}{}_e.\label{casim2}
\eea
Generically these need not be linearly dependent. Since there can be just one independent Killing form in any simple Lie algebra, we would then have a Lie algebra that is not simple. Indeed this is the case for $SO(4)$ and it was also found to be the case for the Minkowski solutions discussed in \cite{Ho:2008ei,Benvenuti:2008bt,Gomis:2008uv}.

We will also find it conventient to introduce the invariant tensors
\bea
g^{abc,def} &=& f^{abc}{}_g f^{defg}\cr
g^{ab} &=& g^{ac,bd} h_{cd}
\eea
later on, where $h_{ab}$ is the inverse of $h^{ab}$.

\subsection{Casimir operators}
Let us first note that
\bea
(t_{ab})^{cd} := \kappa^{-1}_{ab,ef} (t^{ef})^{cd} = \kappa^{-1}_{ab,ef} f^{efcd} = \kappa^{-1}_{ab,ef} \kappa^{ef,cd} = \delta_{ab}^{cd}
\eea
We now list the various expressions for the structure constants, obtained by raising and lowering indices by the Killing form $\kappa^{ab,cd}$:
\bea
C^{ab,cd}{}_{ef} &=& 2 f^{ab[c}{}_{[e} \delta^{d]}_{f]}\cr
C^{ab,cd,ef} &=& 2 f^{ab[c}{}_g f^{d]efg}\cr
C_{ab,cd}{}^{ef} &=& 2 \delta^{[e}_{[a} \delta^{f]}_{{}_[c} h_{b]d_]}
\eea
The last form is nothing but the structure constants of $SO(N)$ if the index range is $a=1,...,N$. Hence $t_{ab}$ generate $SO(N)$. But since $\kappa^{ab,cd}$ need not be the only Killing form, this does not imply that the Lie algebra generated by $t^{ab}$ is also $SO(N)$. 

We will now swith notation and use indices $A, B, ...$ in place of double indices $ab$, $cd$ ,... . So we will for instance denote $g^{ab,cd}$ as $g^{AB}$. Sometimes tensor indices in the Lie 3-algebra are not written out, so for instance $g^{ab}$ and $h^{ab}$ are written just as $g$ and $h$. 

Either form of the structure constants can now be used to show that
\bea
C^{AD}{}_E C^{BE}{}_D = C^{ADE} C_{ED}{}^B =...= 2 g^{AB}
\eea
To show this relation for some certain placements of indices, one may need to use the identity (which follows from the fundamental identity)
\bea
g^{ab,cd} = 2 g^{c[b,a]d}.
\eea

As our next Casimir operators, we have
\bea
g^{AB} t_A t_B &=& g\\
\kappa^{AB} t_A t_B &=& 0\label{caszero}
\eea
where we have suppressed the indices $a,b,...$ on $g^{ab}$ and so on. Here
\bea
g^{AB} &:=& \Tr(t^A t^B)
\eea
where the trace is in the representation specified by Eq (\ref{casim2}).

The group theory factors that arise at one and two-loop are
\bea
C^{ABC} t_A t_B t_C &=& \frac{1}{2} C^{AB}{}_C [t_A,t_B] t^C \cr
&=& \frac{1}{2} C^{AB}{}_C C_{AB}{}^D t_D t^C\cr
&=& \frac{1}{2} 2 g^{DC} t_D t_C\cr
&=& g
\eea
and
\bea
t^A t^B t_A t_B &=& [t^A,t^B] t_A t_B + (t^A t_A)^2\cr
&=& C^{ABC} t_A t_B t_C.\label{casrel}
\eea
We will also need the result
\bea
\kappa_{AC}g^{CB} = \kappa^{AB}
\eea
which may seem confusing as index $A$ is down-stairs in the left hand side, but is up-stairs in the right hand side.

\section{The Bagger-Lambert Lagrangian}\label{action}

We will work in Minkowski signature $(-++)$ on M2 as it is not clear to us how supersymmetry is implemented in Euclidean signature. We then Wick rotate when we compute the loop integrals.

We use eleven-dimensional spinor notation where we make the split $\Gamma^{\mu},\Gamma^I$ associated to $SO(1,2)\times SO(8)$. We define an $SO(8)$ chirality matrix
\bea
\Gamma = \Gamma_{012}
\eea
with the properties 
\bea
\Gamma^2 &=& 1\cr
[\Gamma^{\mu},\Gamma] &=& 0\cr
\{\Gamma^{I},\Gamma\} &=& 0.
\eea
Charge conjugation matrix $C$ is eleven dimensional and subject to $C^T = -C$. It may be defined such that
\bea
(\Gamma^M)^T = -C \Gamma^{M} C^{-1}
\eea
which implies symmetric gamma matrices
\bea
(C \Gamma^M)^T = C\Gamma^M
\eea 
where $M=(\mu,I)$. Supersymmetry parameters are chiral 
\bea
\Gamma \epsilon &=& \epsilon
\eea
and fermions in the theory have opposite chirality,
\bea
\Gamma \psi &=& -\psi.
\eea
We have the duality relation $\Gamma^{\mu\nu} = \epsilon^{\mu\nu\lambda} \Gamma_{\lambda} \Gamma$ where $\epsilon^{012} = -1$.

We denote the fields in the theory as
\bea
X^I &=& X^I_a T^a\cr
\psi &=& \psi_a T^a\cr
A_{\mu} &=& A_{\mu,ab} t^{ab}
\eea
Sometimes we use a short notation $A$ for double indices $ab$ so that we write the gauge field as $A_{\mu,A} t^A$. The gauge covariant derivative acts as
\bea
D_{\mu}X_a &=& \partial_{\mu}X_a - A_{\mu,cd}f^{cdb}{}_a X_b\cr
&=& \partial_{\mu}X_a + A_{\mu,A} (t^A)_{a}{}^b X_b
\eea
The Bagger-Lambert Lagrangian reads \cite{Bagger:2007jr}
\bea
{\cal{L}} &=& \frac{1}{g^2}\Bigg\{-\frac{1}{2}D^{\mu}X^{aI}D_{\mu}X_{a}^I + \frac{i}{2}\bar{\psi}^a\Gamma^{\mu}D_{\mu}\psi_a + \frac{i}{4}f^{abcd}\bar{\psi}_b\Gamma_{IJ}X^I_c X^J_d \psi_a\cr
&&-\frac{1}{12} [X^I,X^J,X^K]^a[X^I,X^J,X^K]_a\cr
&&+\frac{1}{2}\epsilon^{\mu\nu\lambda}\(f^{abcd}A_{\mu ab}\partial_{\nu}A_{\lambda cd} + \frac{2}{3} f^{abc}{}_g f^{defg}A_{\mu ab}A_{\nu cd}A_{\lambda ef}\)\Bigg\}
\eea
where indices $a,b,...$ are contracted using $h^{ab}$. Now recalling that $
f^{abcd} = \kappa^{ab,cd}$ is a Killing form on a Lie algebra with structure constants $C^{ab,cd,ef} = 2f^{ab[c}{}_{g} f^{d]efg}$, we see that the Chern-Simons term can be written as
\bea
\frac{1}{2g^2}\epsilon^{\mu\nu\lambda}\(\kappa^{AB}A_{\mu,A}\partial_{\nu}A_{\lambda,B} + \frac{1}{3} C^{ABC} A_{\mu,A}A_{\nu,B}A_{\lambda,C}\)
\eea
which is the usual Chern-Simons term. Quantum consistency requires the action be well-defined modulo $2\pi$. Due to this Chern-Simons term this implies, at least for $SO(4)$ gauge group, an integer quantization of $\sim g^{-2}$.

We can see that $g$ is a coupling constant by rescaling the fields as
\bea
A &=& gA_{new}\cr
X &=& gX_{new}\cr
\psi &=& g\psi_{new}
\eea
We then drop the subscript $new$. Then the Bagger-Lambert Lagrangian may be viewed as a sum of free plus interacting Lagrangians, 
\bea
{\cal{L}}_0 &=& h^{ab}\(-\frac{1}{2}\partial^{\mu}X^{I}_a\partial_{\mu}X_{b}^I + \frac{i}{2}\bar{\psi}_a\Gamma^{\mu}\partial_{\mu}\psi_b\)\cr
&&+\frac{1}{2}\epsilon^{\mu\nu\lambda}\kappa^{AB}A_{\mu A}\partial_{\nu}A_{\lambda B}
\eea
and 
\bea
{\cal{L}}_{int} &=& -g(t^A)^{ab}\partial^{\mu}X^{I}_a A_{\mu A}X_{b}^I -\frac{g}{6}\epsilon^{\mu\nu\lambda}C^{ABC} A_{\mu A}A_{\nu B}A_{\lambda C} + \frac{ig}{2}(t^A)^{ab}\bar{\psi}_a\Gamma^{\mu}A_{\mu A}\psi_b \cr
&&- \frac{ig^2}{4}f^{abcd}X^I_aX^J_b\bar{\psi}_c\Gamma_{IJ}\psi_d   + \frac{g^2}{2}(t^A t^B)^{ab} A_{\mu A}A_{\mu B} X^I_a X^I_b \cr
&&-\frac{g^4}{12} g^{abc,efg} \delta^{IJK}_{LMN} X^I_a X^J_b X^K_c X^L_e X^M_f X^N_g
\eea
where the interactions are governed by a parameter $g$ that we may take to be small.

Gauge fixing requires this action to be supplemented by a gauge fixing plus ghost term
\bea
\L_{g} = -\frac{1}{\alpha} \kappa^{AB} \partial^{\mu} A_{\mu A}\partial^{\nu} A_{\nu B} + \kappa^{AB} \partial^{\mu}\bar{c}_A D_{\mu}c_B.
\eea
Here the gauge covariant derivative acts on the adjoint index of the ghost field as
\bea
D_{\mu}c_A &=& \partial_{\mu}c_A + A_{\mu B} c_C C^{BC}{}_A
\eea
where the Lie algebra generators are $(t^A)^{B}{}_C = -C^{AB}{}_C$ in the adjoint representation. We can absorb any factor in front of the ghost action by rescaling the ghost fields. Any such rescaling will not affect the covariant derivative so such a factor does not work like a coupling constant -- it is completely irrelevant. The conversion to Bagger-Lambert notation goes as follows,
\bea
D_{\mu} c_{ab} &=& \partial_{\mu}c_{ab} + A_{\mu, cd} c_{ef} C^{cd,ef}{}_{ab}\cr
&=& \partial_{\mu}c_{ab} - (\tilde{A}_{\mu})^e{}_a c_{eb} + (\tilde{A}_{\mu})^{e}{}_{b} c_{ea}
\eea
where $(\tilde{A}_{\mu})^{a}{}_b = A_{\mu,cd} f^{cda}{}_b$.

\section{Renormalization and regularizations}
We write the bare Lagrangian schematically as
\bea
&&A_0 dA_0+g_0{A_0}^3+(dX_0)^2+g_0A_0X_0dX_0 + \psi_0 d\psi_0 + g_0A_0\psi_0\psi_0 \cr
&&+{g_0}^2{A_0}^2{X_0}^2+ {g_0}^2{\psi_0}^2{X_0}^2 +{g_0}^4{X_0}^6+ghosts
\eea
(We have not included calculations of ghost contributions as that would repeat calculations done in pure Chern-Simons theory \cite{Chen:1992ee}.) We then renormalize the bare fields 
\bea
A_0 &=& \sqrt{Z_A} A\cr
X_0 &=& \sqrt{Z_X} X\cr
\psi_0 &=& \sqrt{Z_{\psi}} \psi
\eea
and get
\bea
&& Z_A AdA + g Z_g A^3 + Z_X (dX)^2 + g Z^{(1)}_g AXdX + Z_{\psi} \psi d\psi + g Z^{(2)}_g A\psi\psi \cr
&&g^2 Z^{(3)}_g A^2X^2+ g^2 Z^{(4)}_g \psi^2X^2 + g^4 Z^{(5)}_g X^6+ghosts
\eea
where
\bea
g Z_g &=& g_0{Z_A}^{3/2}\cr
g Z_g^{(1)} &=& g_0 {Z_A}^{1/2}Z_X\cr
g Z_g^{(2)} &=& g_0 {Z_A}^{1/2}Z_{\psi}\cr
g^2 Z_g^{(3)} &=& {g_0}^2{Z_A}Z_X\cr
g^2 Z_g^{(4)} &=& {g_0}^2 Z_{\psi}Z_{X}\cr
g^4 Z_g^{(5)} &=& {g_0}^4 Z_X^3
\eea

The wave function renormalizations $Z_g^{(i)}$ and self-energies $Z_A$, $Z_X$, $Z_{\psi}$ can be obtained for small coupling constant by computing loop diagrams. The one-loop diagrams turn out to be finite and one could think we would not have to care about regularizations if we just compute up to one-loop. However this may not be true. Higher loop diagrams will diverge and so we need to specify some kind regularization. Such a regularization may also affect the finite one-loop diagrams. In this paper we will assume that we use just dimensional regularization. However, this regularization is not obviously gauge invariant when it comes to Chern-Simons theory \cite{Chen:1992ee}. The usual Yang-Mills action is always of the form $F^2$ in any dimension $D$, but the Chern-Simons action is very different in different dimensions and it is not clear apriori that dimensional regularization would preserve gauge symmetry. The source of these problems is associated with the $\epsilon_{\mu\nu\lambda}$ tensor which is difficult to continue analytically to $D$ dimensions, in contrast to eg. the metric tensor which is of the same form in any dimension. One way of performing dimensional regularization in Chern-Simons theory is by putting $k^{\mu}=0$ for all $\mu$-directions corresponding to dimensions above some $D<3$, and then continue the dimension $D$ analytically to make the loop momentum integrals converge, but use the Chern-Simons action in three dimensions.\footnote{This method was suggested to me by Per Salomonson.} 

A better regularization for Chern-Simons theory appears to be to add a Yang-Mills term $-\frac{1}{e^2} F^2$. Since $e^2$ gets a mass dimension in three dimensions it may work as an ultraviolet cutoff. This does not remove all divergencies, but it appears to remove those divergences associated with the $\epsilon_{\mu\nu\lambda}$ tensor \cite{Chen:1992ee}. For that reason one could think that dimensional regularization could always respect gauge symmetry once this Yang-Mills term is included. Unfortunately this Yang-Mills term makes the Feynman rules lot more complicated with a modified gluon propagator and modified gluon vertices, and one also gets a new four-gluon vertex. We have not attempted to compute those much more complicated loop momentum integrals that one gets using such a Yang-Mills term regulator. Dimensional regularization does not violate gauge invariance at one-loop order and therefore we need not use any more sophisticated regularization method here.

\section{Feynman rules}\label{diagram}
Assuming no Yang-Mills regulator term being added, we begin by computing the gluon propagator. In momentum space the crucial term in the Lagrangian is
\bea
\int_p \int_q A_{\mu A}(p) K^{\mu A,\nu B}(p,q) A_{\nu B}(q) 
\eea
with 
\bea
K^{\mu A,\nu B}(p,q) &=& \frac{i}{2}\kappa^{AB}\(\epsilon^{\mu\nu\lambda} p_{\lambda} + \frac{1}{\alpha} p^{\mu}p^{\nu}\)\delta_{pq}
\eea
Here $\int_p := \int \frac{d^3 p}{2\pi}$. The propagator is then given by 
\bea
\frac{i}{2}K^{-1} &=& -\kappa_{AB}\(\epsilon_{\mu\nu\lambda}\frac{p^{\lambda}}{p^2} + \alpha \frac{p_{\mu}p_{\nu}}{p^4}\)
\eea
where $\kappa_{AB}$ is the inverse of $\kappa^{AB}$. In this paper we will choose Landau gauge $\alpha=0$. It is important to note that it is $\kappa_{AB}$ and not $g_{AB}$ or any other Killing form on the non-simple Lie algebra that enters the gluon propagator. This has as a consequence that we only need to rise and lower indices $A, B,...$ using $\kappa_{AB}$. We can never use $g^{AB}$ or any other Killing form in any Feynman diagram exression to contract two adjoint indices $A$ and $B$.

Similar computations give the result that we summarize in the Feynman graphs:

gluon,
\begin{fmffile}{g1}
\begin{fmfgraph*}(20,10)
\fmfleft{i}
\fmfright{o}
\fmf{wiggly}{i,o}
\end{fmfgraph*}
\end{fmffile}
$-\kappa_{AB}\epsilon_{\mu\nu\lambda}\frac{p^{\lambda}}{p^2}$ where the momentum is directed from $\mu$ to $\nu$ 

ghost, 
\begin{fmffile}{g212}
\begin{fmfgraph*}(20,10)
\fmfleft{i}
\fmfright{o}
\fmf{dots}{i,o}
\end{fmfgraph*}
\end{fmffile}
$i\kappa_{AB}\frac{1}{p^2}$

scalar field,
\begin{fmffile}{g3}
\begin{fmfgraph}(20,10)
\fmfleft{i}
\fmfright{o}
\fmf{plain}{i,o}
\end{fmfgraph}
\end{fmffile}
$-i\delta_{ab}\delta^{IJ}\frac{1}{p^2}$

fermion,
\begin{fmffile}{g4}
\begin{fmfgraph}(20,10)
\fmfleft{i}
\fmfright{o}
\fmf{dashes}{i,o}
\end{fmfgraph}
\end{fmffile}
$-i\delta_{ab}\frac{\Gamma^{\mu}_{\alpha\beta} p_{\mu}}{p^2}$ where the momentum is directed from $\alpha$ to $\beta$.

>From the interacting Lagrangian we read off the vertices. Momenta are always directed towards the vertex;

\begin{fmffile}{v1ppp}
\begin{fmfgraph}(20,20)
\fmfleft{i1,i2}
\fmfright{o}
\fmf{wiggly}{i1,v1}
\fmf{wiggly}{i2,v1}
\fmf{wiggly}{v1,o}
\fmfv{label=g}{v1}
\end{fmfgraph}
\end{fmffile}
$-i g C^{ABC} \epsilon_{\mu\nu\lambda}$
\quad
\begin{fmffile}{v1ghost}
\begin{fmfgraph}(20,20)
\fmfleft{i1,i2}
\fmfright{o}
\fmf{dots}{i1,v1}
\fmf{dots}{i2,v1}
\fmf{wiggly}{v1,o}
\fmfv{label=g}{v1}
\end{fmfgraph}
\end{fmffile}
$g C^{ABC} p_{\mu}$

\begin{fmffile}{v1pp}
\begin{fmfgraph}(20,20)
\fmfleft{i1,i2}
\fmfright{o}
\fmf{plain}{i1,v1}
\fmf{plain}{i2,v1}
\fmf{wiggly}{v1,o}
\fmfv{label=g}{v1}
\end{fmfgraph}
\end{fmffile}
$-g (t^A)^{ab}(p_{(a)}-p_{(b)})_{\mu}$
\quad
\begin{fmffile}{v2}
\begin{fmfgraph}(20,20)
\fmfleft{i1,i2}
\fmfright{o}
\fmf{dashes}{i1,v}
\fmf{dashes}{i2,v}
\fmf{wiggly}{v,o}
\end{fmfgraph}
\end{fmffile}
$-g (t^A)^{ab}\Gamma^{\mu}$

\begin{fmffile}{v3}
\begin{fmfgraph}(20,20)
\fmfleft{i1,i2}
\fmfright{o1,o2}
\fmf{plain}{i1,v}
\fmf{plain}{i2,v}
\fmf{wiggly}{v,o1}
\fmf{wiggly}{v,o2}
\end{fmfgraph}
\end{fmffile}
$ig^2 (t^{A} t^{B} + t^Bt^A)^{ab}\delta^{IJ}$
\quad
\begin{fmffile}{v4}
\begin{fmfgraph}(20,20)
\fmfleft{i1,i2}
\fmfright{o1,o2}
\fmf{plain}{i1,v}
\fmf{plain}{i2,v}
\fmf{dashes}{v,o1}
\fmf{dashes}{v,o2}
\end{fmfgraph}
\end{fmffile}
$g^2 f^{abcd}\Gamma_{IJ}$

\begin{fmffile}{v5}
\begin{fmfgraph}(20,20)
\fmfleft{i1,i2,i3}
\fmfright{o1,o2,o3}
\fmf{plain}{i1,v}
\fmf{plain}{i2,v}
\fmf{plain}{i3,v}
\fmf{plain}{v,o1}
\fmf{plain}{v,o2}
\fmf{plain}{v,o3}
\end{fmfgraph}
\end{fmffile}
$-i\frac{g^4}{12}\delta^{IJK}_{LMN} g^{abc,efg}$ plus symmetrized terms in $(Ia,Jb,...)$.

\section{The gluon self-energy}
The one-particle irreducible gluon self-energy may receive quantum corrections and become
\bea
\Pi^{\mu\nu} &=& \Pi_0 \epsilon^{\mu\nu\lambda} \frac{p_{\lambda}}{p^2} + \Pi_e(p)\(\eta^{\mu\nu} - \frac{p^{\mu}p^{\nu}}{p^2}\)
\eea
in the full interacting theory. Let us denote by $\Delta_{\mu\nu}$ the gluon propagator in the interacting theory, and by $\epsilon_{\mu\nu}$ the gluon propagator in the free theory. We get $\Delta_{\mu\nu}$ by summing all one-particle irreducible diagrams, joined by (free theory) gluon propagators. This amounts to the geometric series
\bea
\Delta_{\mu\nu} &=& \epsilon_{\mu\nu} + \epsilon_{\mu\lambda} \Pi^{\lambda\tau} \epsilon_{\tau\nu} +... \cr
&=& \sum_{n=0}^{\infty} \((\epsilon \Pi)^n \epsilon\)_{\mu\nu} = \((1 - \epsilon \Pi)^{-1} \epsilon\)_{\mu\nu}
\eea
We thus need to find the inverse of $1-\epsilon\Pi$. This can be done by making a general ansatz of the same form but with unspecified coefficients. At the end of the day one finds that
\bea
\Delta_{\mu\nu} &=& -\frac{p^2}{p^2 \Pi_e^2 + (p^2-\Pi_0)^2} \(\(p^2-\Pi_0\)\epsilon_{\mu\nu\lambda}\frac{p^{\lambda}}{p^2} + \Pi_e\(\eta_{\mu\nu} - \frac{p_{\mu}p_{\nu}}{p^2}\)\).\label{effective}
\eea

We now compute the one-loop contribution to the self-energy $\Pi^{\mu\nu}$ and begin by computing the diagrams with matter fields in the loop,

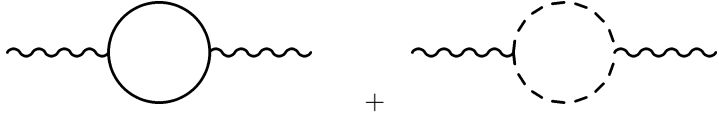
\begin{fmffile}{sub1a}
\begin{fmfgraph}(40,15)
\fmfleft{i}
\fmfright{o}
\fmf{wiggly}{i,v1}
\fmf{plain,left=1,tension=0.5}{v1,v2}
\fmf{plain,left=-1,tension=0.5}{v1,v2}
\fmf{wiggly}{v2,o}
\end{fmfgraph}
\end{fmffile}
\quad +
\begin{fmffile}{sub2aa}
\begin{fmfgraph}(40,15)
\fmfleft{i}
\fmfright{o}
\fmf{wiggly}{i,v1}
\fmf{dashes,left=1,tension=0.5}{v1,v2}
\fmf{dashes,left=-1,tension=0.5}{v1,v2}
\fmf{wiggly}{v2,o}
\end{fmfgraph}
\end{fmffile}
 
The diagram with $X^I$ fields in the loop is
\bea
M_{\mu\nu}^{(X)} = -g^2\Tr(t^A t^B) 8 \int_k \frac{4k_{\mu}k_{\nu}-2\(k_{\mu}p_{\nu}+k_{\nu}p_{\mu}\)+p_{\mu}p_{\nu}}{k^2(p-k)^2}
\eea
Here $8=\delta_I^I$ comes from eight scalar fields going around in the loop.

The diagram with $\psi$ fields in the loop is
\bea
M_{\mu\nu}^{(\psi)} = -g^2\Tr(t^A t^B) \Tr(\Gamma^{\mu}\Gamma^{\sigma}\Gamma^{\nu}\Gamma^{\rho}) \int_k \frac{k_{\rho}(p-k)_{\sigma}}{k^2(p-k)^2}
\eea
where we get an additional minus sign from anticommuting fermions due to the fermion loop. Here we use 
\bea
\Tr(\Gamma^{\mu}\Gamma^{\sigma}\Gamma^{\nu}\Gamma^{\rho}) = 16\(\eta^{\mu\sigma}\eta^{\nu\rho} -\eta^{\mu\nu}\eta^{\sigma\rho} + \eta^{\mu\rho}\eta^{\sigma\nu}\)
\eea
Rather than computing the sum of these two diagrams directly, it can be rewarding to compute each diagram separately in order to check the Ward identity. We use
\bea
\int_k \frac{k_{\mu}k_{\nu}}{k^2(p-k)^2} &=& \frac{|p|}{64}\(3\frac{p_{\mu}p_{\nu}}{p^2}-\eta_{\mu\nu}\)\cr
\int_k \frac{k_{\mu}}{k^2(p-k)^2} &=& \frac{p_{\mu}}{16|p|}\cr
\int_k \frac{1}{k^2(p-k)^2} &=& \frac{1}{8|p|}
\eea
and get
\bea
M_{\mu\nu}^{(X)} = -g^2 g^{AB} \frac{|p|}{2}\(\frac{p_{\mu}p_{\nu}}{|p|^2} - \eta_{\mu\nu}\)\cr
M_{\mu\nu}^{(\psi)} = -g^2 g^{AB} \frac{|p|}{2}\(\frac{p_{\mu}p_{\nu}}{|p|^2} - \eta_{\mu\nu}\).
\eea
The Ward identity $p^{\mu}M_{\mu\nu}=0$ is obeyed for both diagrams separately. One could think the diagrams, being identical in magnitude, would cancel each other in a supersymmetric theory. But that does not happen here. Quite the contrary, they add up.\footnote{This was pointed out to me by Soo-Jong Rey.}

We have only two more diagrams at one-loop contributing to the self-energy. They exactly cancel,

\begin{fmffile}{sub3}
\begin{fmfgraph}(40,15)
\fmfleft{i}
\fmfright{o}
\fmf{wiggly}{i,v1}
\fmf{wiggly,left=1,tension=0.5}{v1,v2}
\fmf{wiggly,left=-1,tension=0.5}{v1,v2}
\fmf{wiggly}{v2,o}
\end{fmfgraph}
\end{fmffile}
\quad +
\begin{fmffile}{sub4}
\begin{fmfgraph}(40,15)
\fmfleft{i}
\fmfright{o}
\fmf{wiggly}{i,v1}
\fmf{dots,left=1,tension=0.5}{v1,v2}
\fmf{dots,left=-1,tension=0.5}{v1,v2}
\fmf{wiggly}{v2,o}
\end{fmfgraph}
\end{fmffile}=0. 
This cancellation was shown in \cite{Chen:1992ee} for pure Chern-Simons theory, and we can use that computation with no modification for any coupling to matter fields.

We note that the diagram 

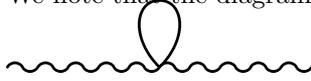
\begin{fmffile}{referee}
\begin{fmfgraph}(40,15)
\fmfleft{i1}
\fmfright{o1}
\fmf{wiggly}{i1,v1,o1}
\fmf{plain,left=-1.2,tension=1.5}{v1,v1}
\end{fmfgraph}
\end{fmffile}
vanishes because 
\bea
\int_k \frac{1}{k^2} = 0
\eea
using dimensional regularization. 

We conclude that the one-loop contribution to the self-energy is given by
\bea
\Pi_{\mu\nu} = -g^2 g^{AB} \frac{1}{|p|}\(p_{\mu}p_{\nu}-p^2\eta_{\mu\nu}\).
\eea
Plugging this into Eq (\ref{effective}) we get the quantum corrected propagator
\bea
\Delta_{\mu\nu} = -\frac{1}{1+g^4}\epsilon_{\mu\nu\lambda}\frac{p^{\lambda}}{p^2} - \frac{g^2}{1+g^4} \frac{1}{|p|}\(\eta_{\mu\nu}-\frac{p_{\mu}p_{\nu}}{p^2}\).
\eea
This in turn can be obtained at tree level from an effective action which contains a kinetic term with kernel\footnote{ignoring the gauge fixing term}
\bea
K^{\mu\nu} = \frac{i}{2}\kappa^{AB}\(\epsilon^{\mu\nu\lambda}p_{\lambda} - g^2\frac{1}{|p|}\(\eta^{\mu\nu}p^2-p^{\mu}p^{\nu}\)\).
\eea
We conclude that 
\bea
Z_A = 1 + \Ordo(g^4)
\eea
there being no one-loop correction to the kinetic term $\frac{i}{2}\kappa^{AB}\epsilon^{\mu\nu\lambda}p_{\lambda}$. However, we find a new term in the effective action that is a non-local term, of the form
\bea
g^2\int d^3p \frac{1}{|p|}\(\eta^{\mu\nu}p^2-p^{\mu}p^{\nu}\)A_{\mu}(p)A_{\nu}(-p).
\eea
In position space we find that this term is given by a non-local expression as can be seen by computing the fourier transform (using eg the technique of fractional derivatives)
\bea
\int d^3 p \sqrt{p^2} e^{ip.(x-y)}\sim \frac{1}{|x-y|^4}.
\eea

\section{Scalar field self-energy}
We find that all one-loop corrections to the self-energy of the scalar fields are exactly zero, for reasons as indicated in the diagrams below,

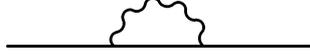
\begin{fmffile}{one-loop1aa}
\begin{fmfgraph}(40,15)
\fmfleft{i1}
\fmfright{o1}
\fmf{plain}{i1,v1,v2,o1}
\fmf{wiggly,left=1,tension=0.2}{v1,v2}
\end{fmfgraph}
\end{fmffile}
\bea
\sim \epsilon_{\mu\nu\lambda}(p+k)^{\mu}(p+k)^{\nu}(p-k)^{\lambda} \equiv 0
\eea

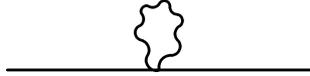
\begin{fmffile}{one-loop2aabbbb}
\begin{fmfgraph}(40,15)
\fmfleft{i1}
\fmfright{o1}
\fmf{plain}{i1,v1,o1}
\fmf{wiggly,left=-1.2,tension=1.5}{v1,v1}
\end{fmfgraph}
\end{fmffile}
\bea
\sim \eta^{\mu\nu}\epsilon_{\mu\nu\lambda}\frac{k^{\lambda}}{k^2} \equiv 0
\eea

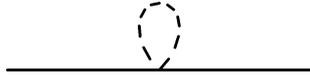
\begin{fmffile}{one-loop2aabbbbc}
\begin{fmfgraph}(40,15)
\fmfleft{i1}
\fmfright{o1}
\fmf{plain}{i1,v1,o1}
\fmf{dashes,left=-1.2,tension=1.5}{v1,v1}
\end{fmfgraph}
\end{fmffile}
\bea
\sim f^{abcd}\delta_{cd} \equiv 0
\eea

\section{$X^6$ corrections}
From the renormalized action we read off the corrected six-point vertex $\sim g^4 Z_g^{(5)}$. According to the relation $g^4 Z_g^{(5)} = {g_0}^4 Z_X^3$, non-renormalization of the coupling constant means that we should have $Z_g^{(5)}=1 + \Ordo(g^4)$ as we in the previous section found no one-loop corrections to the scalar field propagator, i.e. $Z_X=1+\Ordo(g^4)$. There are $6$ different types of one-loop diagrams that contribute to $Z_g^{(5)}$,

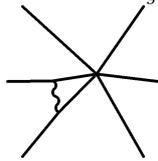
\begin{fmffile}{6p1}
\begin{fmfgraph}(20,20)
\fmfleft{i1,i2,i3}
\fmfright{o1,o2,o3}
\fmf{plain}{i1,v1,v}
\fmf{plain}{i2,v2,v}
\fmf{plain}{i3,v}
\fmf{plain}{v,o1}
\fmf{plain}{v,o2}
\fmf{plain}{v,o3}
\fmf{wiggly,tension=0.2}{v1,v2}
\end{fmfgraph}
\end{fmffile}
Group theory factor:
\bea
\kappa_{AB}(t^A)^{am}(t^B)^{bn} \(g^{mnc,def} \delta^{IJK}_{LMN} + symm\)
\eea
where symm means symmetrized in $mI, nJ,cK,..., fN$. We should then sum over all diagrams obtained by permuting $aI,bJ,cK,...,fN$.

We now use that
\bea
\kappa_{AB}(t^A)^{am} (t^B)^{bn} &=& f^{ambn}
\eea
and find that the group theory factor can be rewritten as
\bea
f^{mn[ab} g^{cde,f]mn} &=& f^{mn[ab} f^{cde}{}_g f^{f]mng}
\eea
This now vanishes by the fundamental identity in the form
\bea
f^{[cde}{}_g f^{f]mng} = 0.
\eea
This we consider as our main result in this paper. The emergence of the fundamental identity in this loop diagram makes us believe that this will always work like this for any loop diagram. We think that the fundamental identity is precisely what is needed in order for the coupling constant to not renormalize.

It remains to analyze the other one-loop diagram corrections to the six-point vertex, but we will not find the fundamental identity for these diagrams, but they rather cancel or vanish for other reasons,

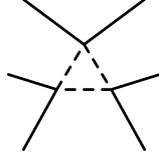
\begin{fmffile}{6p22}
\begin{fmfgraph}(20,20)
\fmfleft{i1,i2,i3}
\fmfright{o1,o2,o3}
\fmf{plain}{i1,v1}
\fmf{plain}{i2,v1}
\fmf{plain}{i3,v2}
\fmf{plain}{v2,o3}
\fmf{plain}{v3,o2}
\fmf{plain}{v3,o1}
\fmf{dashes}{v1,v2}
\fmf{dashes}{v2,v3}
\fmf{dashes}{v3,v1}
\end{fmfgraph}
\end{fmffile}
For a particular choice of external legs labeled by $Ia, Jb, Kc, Ld, Me, Nf$ we get
\bea
&&ig^6 \Tr(t^{ab}t^{cd}t^{ef})\Tr(\Gamma_{IJ}\Gamma^{\mu}\Gamma_{KL}\Gamma^{\nu}\Gamma_{MN}\Gamma^{\lambda})\cr
&&\times \int_k \frac{k_{\mu}(k+q)_{\nu}(k-p)_{\lambda}}{k^2(k+q)^2(k-p)^2}
\eea
We then note\footnote{Here we trace only over the anti-chiral parts, that is, we should really insert a projector $P=\frac{1}{2}(1-\Gamma)$ inside the trace. Then $Tr(\Gamma^{\mu\nu\lambda}P) = \epsilon^{\mu\nu\lambda}\Tr(\Gamma P) = -\frac{1}{2} \epsilon^{\mu\nu\lambda} Tr(\Gamma^2) = -\frac{32}{2} \epsilon^{\mu\nu\lambda}$.} 
\bea
\Tr(\Gamma_{IJ}\Gamma^{\mu}\Gamma_{KL}\Gamma^{\nu}\Gamma_{MN}\Gamma^{\lambda}) &=& -8\delta_{[I}^{[K} \delta_{J][M} \delta^{L]}_{N]} \epsilon^{\mu\nu\lambda}
\eea
We should sum all diagrams that we obtain by permuting the labels $Ia,JB,...$. Gathering all the term associated with the factor $\delta_{IJ}\delta_{KL}\delta_{MN}$ we find 
\bea
-\Tr(t^{ac}t^{be}t^{df}) + permutations
\eea
where permutations amounts to antisymmetrization each of the pairs $ab$, $cd$ and $ef$ respectively, which produces $8$ terms.

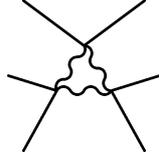
\begin{fmffile}{6p3a}
\begin{fmfgraph}(20,20)
\fmfleft{i1,i2,i3}
\fmfright{o1,o2,o3}
\fmf{plain}{i1,v1}
\fmf{plain}{i2,v1}
\fmf{plain}{i3,v2}
\fmf{plain}{v2,o3}
\fmf{plain}{v3,o2}
\fmf{plain}{v3,o1}
\fmf{wiggly}{v1,v2}
\fmf{wiggly}{v2,v3}
\fmf{wiggly}{v3,v1}
\end{fmfgraph}
\end{fmffile}
Again we give the expression for one particular choice of labeling of the external legs,
\bea
&&ig^6 \delta_{IJ}\delta_{KL}\delta_{MN} \(\Tr(t^{ac}t^{be}t^{df}) + permutations\)\cr
&&\times\epsilon^{\mu\nu\lambda}\int_k \frac{k_{\mu}(k+q)_{\nu}(k-p)_{\lambda}}{k^2(k+q)^2(k-p)^2}
\eea
and we see that this cancels the correspondig term coming from the sum of the diagrams above with fermions running in the loop.

The remaining one-loop diagrams are identically zero, and this has kinematic reasons. The vanishing of these diagrams is therefore independent of the choice of Lie 3-algebra,

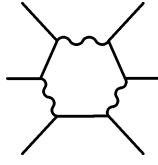
\begin{fmffile}{6p3a1}
\begin{fmfgraph}(20,20)
\fmfleft{i1,i2,i3}
\fmfright{o1,o2,o3}
\fmf{plain}{i1,v1}
\fmf{plain}{i2,v2}
\fmf{plain}{i3,v3}
\fmf{plain}{v4,o3}
\fmf{plain}{v5,o2}
\fmf{plain}{v6,o1}
\fmf{wiggly}{v1,v2}
\fmf{wiggly}{v3,v4}
\fmf{wiggly}{v5,v6}
\fmf{plain}{v2,v3}
\fmf{plain}{v4,v5}
\fmf{plain}{v6,v1}
\end{fmfgraph}
\end{fmffile}
Letting the five independent ingoing momenta be denoted as $p_1,...,p_5$ and the loop momentum $k$, we find
\bea
&\sim &(p_1+k)^{\mu}(p_1+2p_2-k)^{\nu}(p_3-p_1-p_2-k)^{\kappa}\cr
&&(p_1+p_2+p_3+2p_4-l)^{\tau}(p_1+p_2+p_3+p_4+p_5-k)^{\rho}(p_1+p_2+p_3+p_4+p_5+k)^{\sigma}\cr
&&\epsilon_{\mu\nu\alpha}(p-k)^{\alpha}\epsilon_{\tau\kappa\beta}(k-p_1-p_2-p_3)^{\beta}\epsilon_{\sigma\rho\gamma}(k-p_1-p_2-p_3-p_4-p_5)^{\gamma}\cr
&\equiv & 0
\eea
(this is seen by using the complete antisymmetry of $\epsilon_{\sigma\rho\gamma}$.)

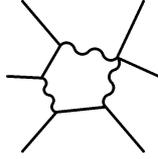
\begin{fmffile}{6p3a2}
\begin{fmfgraph}(20,20)
\fmfleft{i1,i2,i3}
\fmfright{o1,o2,o3}
\fmf{plain}{i1,v1}
\fmf{plain}{i2,v2}
\fmf{plain}{i3,v3}
\fmf{plain}{v4,o3}
\fmf{plain}{v4,o2}
\fmf{plain}{v6,o1}
\fmf{wiggly}{v1,v2}
\fmf{wiggly}{v3,v4}
\fmf{wiggly}{v4,v6}
\fmf{plain}{v2,v3}
\fmf{plain}{v6,v1}
\end{fmfgraph}
\end{fmffile}
\bea
\sim \epsilon_{\kappa\tau\gamma}(k+p)^{\tau}(k+p)^{k+p} \equiv 0
\eea

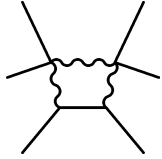
\begin{fmffile}{6p3a3}
\begin{fmfgraph}(20,20)
\fmfleft{i1,i2,i3}
\fmfright{o1,o2,o3}
\fmf{plain}{i1,v1}
\fmf{plain}{i2,v2}
\fmf{plain}{i3,v2}
\fmf{plain}{v3,o3}
\fmf{plain}{v3,o2}
\fmf{plain}{v4,o1}
\fmf{wiggly}{v1,v2}
\fmf{wiggly}{v2,v3}
\fmf{wiggly}{v3,v4}
\fmf{plain}{v4,v1}
\end{fmfgraph}
\end{fmffile}
\bea
\sim \epsilon_{\kappa\tau\gamma}(k+p)^{\tau}(k+p)^{k+p} \equiv 0
\eea

\section{$XXA$ corrections}
We now compute one-loop corrections to the $XXA$ vertex 

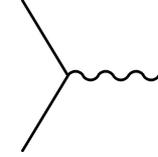
\begin{fmffile}{v1pp}
\begin{fmfgraph}(20,20)
\fmfleft{i1,i2}
\fmfright{o}
\fmf{plain}{i1,v1}
\fmf{plain}{i2,v1}
\fmf{wiggly}{v1,o}
\fmfv{label=g}{v1}
\end{fmfgraph}
\end{fmffile}
$-g (t^A)^{ab}(p_{(a)}-p_{(b)})_{\mu}$. 

Quantum correction generate a new type of $XXA$ vertex, that should be of the form $(t^A)^{ab}\epsilon_{\mu\nu\lambda}p^{\nu}_a p^{\mu}_b$ times some form factor. We will not care much about this contribution as it will not affect the coupling constant of the original vertex, nor does it violate the Ward identity. Instead we will isolate only the contribution that is of the form $(t^A)^{ab}(p_{(a)}-p_{(b)})_{\mu}$ times some form factor.

First we note that the following diagrams are identically zero,

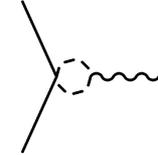
\begin{fmffile}{sv2bx}
\begin{fmfgraph}(20,20)
\fmfleft{i1,i2}
\fmfright{o}
\fmf{plain}{i1,w,i2}
\fmf{wiggly}{v,o}
\fmf{dashes,left=1}{w,v}
\fmf{dashes,left=1}{v,w}
\end{fmfgraph}
\end{fmffile}
$\Tr(\Gamma_{IJ}\Gamma_{\mu}) \equiv 0$,
\quad
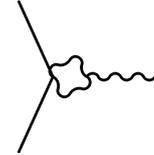
\begin{fmffile}{sv3bx}
\begin{fmfgraph}(20,20)
\fmfleft{i1,i2}
\fmfright{o}
\fmf{plain}{i1,w,i2}
\fmf{wiggly}{v,o}
\fmf{wiggly,left=1}{w,v}
\fmf{wiggly,left=1}{v,w}
\end{fmfgraph}
\end{fmffile}
$C^{ABC} (t_Bt_C+t_Ct_B) \equiv 0$
\newline
and so we are left with only two more diagrams. The first is

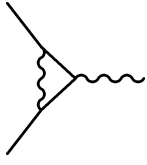
\begin{fmffile}{sv1}
\begin{fmfgraph}(20,20)
\fmfleft{i1,i2}
\fmfright{o}
\fmf{plain}{i1,w,v}
\fmf{plain}{i2,u,v}
\fmf{wiggly}{v,o}
\fmf{wiggly,tension=0.2}{u,w}
\end{fmfgraph}
\end{fmffile}
\bea
=g^3 t^B t^A t_B \int_k \frac{4\epsilon_{\nu\rho\lambda}p_a^{\nu}p_b^{\rho}k^{\lambda}(p_a-p_b-2k)_{\mu}}{k^2(p_a-k)^2(p_b+k)^2}
\eea
We now note that
\bea
\int_k \frac{\epsilon_{\mu\nu\rho} p_a^{\mu} p_b^{\nu} k^{\rho}}{k^2(p_a-k)^2(p_b-k)^2} \equiv 0
\eea
This can be seen with no computations, just by noting that the result one gets by performing the integration the must still contain the $\epsilon_{\mu\nu\rho}$ factor, and there are just two independent momenta $p_a$ and $p_b$ that it can be contracted by. This means we get no correction to the vertex $(t^A)^{ab} (p_a-p_b)_{\mu}$ from this diagram.
 
We have one more diagram,

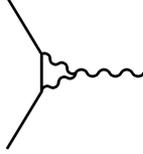
\begin{fmffile}{sv2a}
\begin{fmfgraph}(20,20)
\fmfleft{i1,i2}
\fmfright{o}
\fmf{plain}{i1,w,u,i2}
\fmf{wiggly}{v,o}
\fmf{wiggly}{v,w}
\fmf{wiggly}{v,u}
\end{fmfgraph}
\end{fmffile}
\bea
=-g^3 C^{ABC} t_B t_C \int_k \frac{2\epsilon_{\nu\rho\lambda}p_a^{\nu}p_b^{\rho}k^{\lambda}(p_a-p_a-2k)_{\mu}}{k^2(p_a-k)^2(p_b+k)^2}
\eea

The sum of these two diagrams is
\bea
-g^3\(2t^B t^A t_B - C^{ABC} t_{B} t_C\) \int_k \frac{4 \epsilon_{\nu\rho\lambda}p_a^{\nu}p_b^{\rho}k^{\lambda} k_{\mu}}{k^2(p_a-k)^2(p_b+k)^2}
\eea
which is non-vanishing. So we get a new type of $XXA$ vertex in the effective theory. But we get no one-loop correction to the coupling constant associated with the original $XXA$ vertex.

\section{Fermion self-energy}
We first note that the following two diagrams are identically zero,
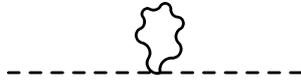
\newline

\begin{fmffile}{f2}
\begin{fmfgraph}(40,15)
\fmfleft{i1}
\fmfright{o1}
\fmf{dashes}{i1,v1,o1}
\fmf{wiggly,left=-1.2,tension=1.5}{v1,v1}
\end{fmfgraph}
\end{fmffile}
\bea
\sim \eta^{\mu\nu}\epsilon_{\mu\nu\lambda}\frac{k^{\lambda}}{k^2} \equiv 0
\eea

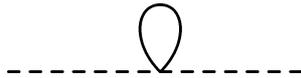
\begin{fmffile}{f3}
\begin{fmfgraph}(40,15)
\fmfleft{i1}
\fmfright{o1}
\fmf{dashes}{i1,v1,o1}
\fmf{plain,left=-1.2,tension=1.5}{v1,v1}
\end{fmfgraph}
\end{fmffile}
\bea
\sim f^{abcd}\delta_{cd} \equiv 0
\eea
\newline
and only the diagram 

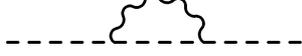
\begin{fmffile}{f1}
\begin{fmfgraph}(40,15)
\fmfleft{i1}
\fmfright{o1}
\fmf{dashes}{i1,v1,v2,o1}
\fmf{wiggly,left=1,tension=0.2}{v1,v2}
\end{fmfgraph}
\end{fmffile}
\newline
is potentially non-vanishing. However the group theory factor associated with this diagram is
\bea
t^A t_A
\eea
which vanishes identically for all Lie 3-algebras by Eq (\ref{caszero}).

\section{$A\psi\psi$ corrections}
First we note that the following diagram is identically zero,

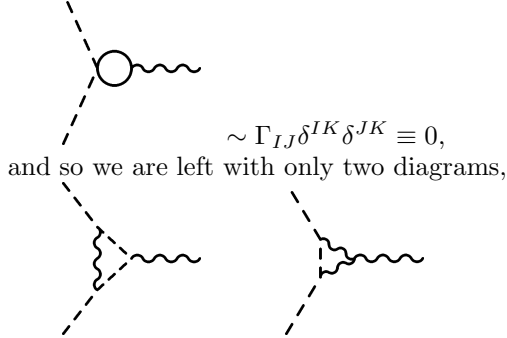
\begin{fmffile}{fv2b}
\begin{fmfgraph}(20,20)
\fmfleft{i1,i2}
\fmfright{o}
\fmf{dashes}{i1,w,i2}
\fmf{wiggly}{v,o}
\fmf{plain,left=1}{w,v}
\fmf{plain,left=1}{v,w}
\end{fmfgraph}
\end{fmffile}
$\sim \Gamma_{IJ}\delta^{IK}\delta^{JK} \equiv 0$, \newline
and so we are left with only two diagrams,

\begin{fmffile}{fv1}
\begin{fmfgraph}(20,20)
\fmfleft{i1,i2}
\fmfright{o}
\fmf{dashes}{i1,w,v}
\fmf{dashes}{i2,u,v}
\fmf{wiggly}{v,o}
\fmf{wiggly,tension=0.2}{u,w}
\end{fmfgraph}
\end{fmffile}
\quad
\begin{fmffile}{fv2ax}
\begin{fmfgraph}(20,20)
\fmfleft{i1,i2}
\fmfright{o}
\fmf{dashes}{i1,w,u,i2}
\fmf{wiggly}{v,o}
\fmf{wiggly}{v,w}
\fmf{wiggly}{v,u}
\end{fmfgraph}
\end{fmffile}
\newline
Let us denote by
\bea
I_{\mu\nu\lambda}(p_a,p_b)&\equiv &\int_k \frac{(p_a - k)_{\mu} (p_b+k)_{\nu} k_{\lambda}}{(p_a-k)^2(p_b+k)^2k^2}
\eea
the momentum integral that occurs in these two diagrams. It can be shown that 
\bea
\epsilon^{\mu\nu\lambda}I_{\mu\nu\lambda}=0
\eea
and this will be the only property we will really need of this integral. Then  
\bea
{\mbox{\begin{fmffile}{fv1}
\begin{fmfgraph}(20,20)
\fmfleft{i1,i2}
\fmfright{o}
\fmf{dashes}{i1,w,v}
\fmf{dashes}{i2,u,v}
\fmf{wiggly}{v,o}
\fmf{wiggly,tension=0.2}{u,w}
\end{fmfgraph}
\end{fmffile}}}
&=&-g^3t_Bt^At^B \Gamma^{\nu}\Gamma^{\alpha}\Gamma^{\mu}\Gamma^{\beta}\Gamma^{\rho}\epsilon_{\nu\rho\lambda} I_{\alpha\beta}{}^{\lambda} \equiv A^A_{\mu}
\eea
and
\bea
\mbox{\begin{fmffile}{fv2a}
\begin{fmfgraph}(20,20)
\fmfleft{i1,i2}
\fmfright{o}
\fmf{dashes}{i1,w,u,i2}
\fmf{wiggly}{v,o}
\fmf{wiggly}{v,w}
\fmf{wiggly}{v,u}
\end{fmfgraph}
\end{fmffile}}
&=&g^3C^{ABC}t_Bt_C \Gamma^{\nu}\Gamma^{\alpha}\Gamma^{\kappa} \epsilon_{\mu\rho\tau}\epsilon_{\kappa\tau\lambda}\epsilon_{\nu\rho\sigma}I^{\sigma\lambda}{}_{\alpha} \equiv B^A_{\mu}
\eea
where we have noted the overall factors of $6.4.2/3! = 8$ and $4.3.2.2/2 = 24$ respectively, from possible contractions. The minus sign in the first diagram is due to fermions anticommuting when we perform the contractions.

We now need the identities
\bea
\Gamma^{\mu}\Gamma^{\alpha}\Gamma^{\beta}\Gamma^{\gamma}\Gamma^{\nu}\epsilon_{\mu\nu\lambda} &=& 2\epsilon^{\alpha\beta\gamma}\Gamma_{\lambda}-2\eta^{\alpha\beta}\delta^{\gamma}_{\lambda}-2\eta^{\beta\gamma}\delta^{\alpha}_{\lambda},\cr
\Gamma^{\nu}\Gamma^{\alpha}\Gamma^{\kappa} \epsilon_{\mu\rho\tau}\epsilon_{\kappa\tau\lambda}\epsilon_{\nu\rho\sigma} &=& -\delta_{\mu}^{\sigma}\delta_{\alpha}^{\lambda}-\delta_{\mu}^{\lambda}\delta_{\alpha}^{\sigma}-\epsilon^{\alpha\lambda\sigma}\Gamma^{\mu} + \epsilon^{\mu\lambda\sigma}\Gamma^{\alpha} + \delta^{\alpha}_{\mu}\epsilon^{\sigma\lambda\kappa}\Gamma_{\kappa}
\eea
We then get
\bea
A^A_{\mu} &=& -2g^3t_Bt^At^B \(-I_{\mu\lambda}{}^{\lambda}-I_{\lambda\mu}{}^{\lambda}-\epsilon^{\mu\alpha\beta}I_{\alpha\beta}{}^{\lambda}\Gamma_{\lambda}\)
\eea
and 
\bea
B^A_{\mu} &=& g^3C^{ABC}t_Bt_C \(-I_{\mu\lambda}{}^{\lambda}-I_{\lambda\mu}{}^{\lambda}-\epsilon^{\mu\alpha\beta}I_{\alpha\beta\lambda}\Gamma^{\lambda}-\epsilon^{\lambda\alpha\beta}I_{\alpha\beta\mu}\Gamma_{\lambda}\)
\eea
respectively. 

We are now only interested in corrections which are of the same form as the original vertex, that is, of the form $(t^A)^{ab}\Gamma^{\mu}$. Hence we isolate the piece of $\epsilon^{\mu\alpha\beta} I_{\alpha\beta\lambda}$ that is proportional to $\delta^{\mu}_{\lambda}$. If we Write $\epsilon^{\mu\alpha\beta} I_{\alpha\beta\lambda} = \delta_{\lambda}^{\mu} I +...$ where the dots could involve terms like $(p_a)_{\lambda} (p_b)_{\mu} J + ...$, then we find the one-loop correction
\bea
2g^3(t_B t^A t^B - C^{ABC}t_B t_C) I \Gamma^{\mu}
\eea
to the original vertex. Now this vanishes only when $t_A t_B t^A t^B = C^{ABC}t_A t_B t_C$, which is the case whenever the fundamental identity is satisfied according to Eq (\ref{casrel}).

We have now demonstrated that the Bagger-Lambert action yields a consistent, i.e. gauge invariant, quantum theory up to one-loop order for any choice of Lie 3-algebra.

Also we have seen that quantum corrections generate new albeit non-local terms in the effective action. It could be interesting to investigate these terms more carefully and for instance examine how supersymmetry works when these non-local terms are included in the action.
 
\vskip20pt
{\sl{Acknowledgements.}} I have benefitted from discussions with U. Gran and B.E.W Nilsson, and in particular P. Salomonson.


\newpage


\begin{thebibliography}{999}
\bibitem{Witten:1988hf}
  E.~Witten,
  Commun.\ Math.\ Phys.\  {\bf 121} (1989) 351.

\bibitem{Chen:1992ee}
  W.~Chen, G.~W.~Semenoff and Y.~S.~Wu,
  Phys.\ Rev.\  D {\bf 46} (1992) 5521
  [arXiv:hep-th/9209005].


\bibitem{Bagger:2007vi}
  J.~Bagger and N.~Lambert,
  JHEP {\bf 0802} (2008) 105
  [arXiv:0712.3738 [hep-th]].

\bibitem{Bagger:2007jr}
  J.~Bagger and N.~Lambert,
  Phys.\ Rev.\  D {\bf 77} (2008) 065008
  [arXiv:0711.0955 [hep-th]].

\bibitem{Gustavsson:2007vu}
  A.~Gustavsson,
  arXiv:0709.1260 [hep-th].

\bibitem{Gustavsson:2008dy}
  A.~Gustavsson,
  JHEP {\bf 0804} (2008) 083
  [arXiv:0802.3456 [hep-th]].

\bibitem{Ho:2008ei}
  P.~M.~Ho, Y.~Imamura and Y.~Matsuo,
  arXiv:0805.1202 [hep-th].
\bibitem{Benvenuti:2008bt}
  S.~Benvenuti, D.~Rodriguez-Gomez, E.~Tonni and H.~Verlinde,
  arXiv:0805.1087 [hep-th].
\bibitem{Gomis:2008uv}
  J.~Gomis, G.~Milanesi and J.~G.~Russo,
  arXiv:0805.1012 [hep-th].
\bibitem{Lin:2008qp}
  H.~Lin,
  arXiv:0805.4003 [hep-th].


\bibitem{Ho:2008nn}
  P.~M.~Ho and Y.~Matsuo,
  arXiv:0804.3629 [hep-th].

\bibitem{Papadopoulos:2008gh}
  G.~Papadopoulos,
  arXiv:0804.3567 [hep-th].
\bibitem{Gauntlett:2008uf}
  J.~P.~Gauntlett and J.~B.~Gutowski,
  arXiv:0804.3078 [hep-th].


\bibitem{Gran:2008vi}
  U.~Gran, B.~E.~W.~Nilsson and C.~Petersson,
  arXiv:0804.1784 [hep-th].

\bibitem{Bandres:2008vf}
  M.~A.~Bandres, A.~E.~Lipstein and J.~H.~Schwarz,
  JHEP {\bf 0805} (2008) 025
  [arXiv:0803.3242 [hep-th]].

\bibitem{FigueroaO'Farrill:2008zm}
  J.~Figueroa-O'Farrill, P.~de Medeiros and E.~Mendez-Escobar,
  arXiv:0805.4363 [hep-th].

\bibitem{Kapustin:1994mt}
  A.~N.~Kapustin and P.~I.~Pronin,
  Mod.\ Phys.\ Lett.\  A {\bf 9}, 1925 (1994)
  [arXiv:hep-th/9401053].

\bibitem{Gaiotto:2007qi}
  D.~Gaiotto and X.~Yin,
  JHEP {\bf 0708}, 056 (2007)
  [arXiv:0704.3740 [hep-th]].

\bibitem{Aharony:2008ug}
  O.~Aharony, O.~Bergman, D.~L.~Jafferis and J.~Maldacena,
  arXiv:0806.1218 [hep-th].


\end{thebibliography}
 \end{document}